\documentstyle[twocolumn,aps,psfig]{revtex}

\begin{document}
\draft
\flushbottom
\twocolumn[
\hsize\textwidth\columnwidth\hsize\csname @twocolumnfalse\endcsname

\title{Anomalous field-dependent specific heat in charge-ordered Pr$_{1-x}$Ca$_x$MnO%
$_3$ and La$_{0.5}$Ca$_{0.5}$MnO$_3$}
\author{V. N. Smolyaninova$^1$, Amlan Biswas$^1$, X. Zhang$^1$, K. H. Kim$^2$,
Bog-Gi~Kim$^2$, S-W.~Cheong$^2$, and R. L. Greene$^1$}
\address{1. Department of Physics and Center for Superconductivity Research,\\
University of Maryland, College Park,\\
MD 20742}
\address{2. Department of Physics and Astronomy, Rutgers University, Piscatway,\\
New Jersey 08854}
\date{\today}
\maketitle
\tightenlines
\widetext
\advance\leftskip by 57pt
\advance\rightskip by 57pt

\begin{abstract}
We report low temperature specific heat measurements of
Pr$_{1-x}$Ca$_{x}$MnO$_{3}$  ($0.3\leq x \leq 0.5$)
and La$_{0.5}$Ca$_{0.5}$MnO$_{3}$ with and
without applied magnetic field.
An excess specific heat, $C^{\prime}(T)$, of non-magnetic
origin associated with charge ordering is found for all the samples. A
magnetic field sufficient to induce the transition from the
charge-ordered state to the ferromagnetic metallic state
does not completely remove  the $C^{\prime }$ contribution. This
suggests that the charge ordering is not completely destroyed by a
``melting" magnetic
field. In addition, the
specific heat of the Pr$_{1-x}$Ca$_{x}$MnO$_{3}$ compounds exhibit a large
contribution linear in
temperature ($\gamma T$) originating from magnetic and charge disorder.


PACS number(s): 75.40.Cx, 75.30.Vn, 71.30.+h, 75.50.Cc
\end{abstract}

\pacs{PACS no.: 75.40.Cx, 75.30.Vn, 71.30.+h, 75.50.Cc}

]
\narrowtext
\tightenlines


Charge ordering (CO), i.e. the real space ordering of Mn$^{3+}$ and Mn$^{4+}$
ions, is one of the most intriguing properties observed in hole-doped
manganites. These compounds with the generalized formula RE$_{1-x}$AE$_{x}$%
MnO$_{3}$ (RE being a trivalent rare-earth and AE being a divalent alkaline
earth element occupying the A-site in the AMnO$_{3}$ perovskite structure)
undergo a charge-ordering transition for certain values of $x$ and the
average A-site cation radius $<r_{A}>$. Remarkably, a modest external
magnetic field can destroy the insulating CO state and produce a metallic
ferromagnetic state (FMM). The $<r_{A}>$ determines the one-electron
bandwidth $W$ in these materials and charge ordering is observed in
materials with $x=0.5$ with small $<r_{A}>\sim 1.23$ \AA ~ (and consequently
small $W$).

Among these, Pr$_{1-x}$Ca$_{x}$MnO$_{3}$ ($<r_{A}>\sim 1.18$ \AA ~) is
especially interesting. This compound is a paramagnetic insulator at high
temperature which undergoes a
CO transition at $T_{{\rm CO}}$ $\sim 230$ K for the composition range $%
0.3\le x\le 0.5$. An antiferromagnetic (AFM) ordering occurs below $T_{{\rm %
CO}}$ with Neel temperature ($T_{N}$) changing from $\sim 180$ K ($x=0.5$)
to $\sim 140$ K ($x=0.3$) \cite{Jirak,Yoshizawa,Tomioka}. The AFM ordering
for charge-ordered Pr$_{1-x}$Ca$_{x}$MnO$_{3}$ is CE type for $x\approx 0.5$ 
\cite{Jirak} and pseudo-CE type for $x\approx 0.3$ \cite{Jirak,Yoshizawa,Cox}%
. Although the FMM state is never realized for Pr$_{1-x} $Ca$_{x}$MnO$_{3}$
in zero magnetic field, a competition between the ferromagnetic (FM)
metallic and AFM charge-ordered ground states leads to an increase of FM
tendencies as $x$ decreases below 0.5.
The magnitude of the magnetic field required to induce a
transition to the FMM state decreases from 24 T for $x=0.5$ \cite{Tokunaga}
to 4 T for $x=0.3$ \cite{Yoshizawa}. The charge ordering has been detected as a
superlattice reflection in synchrotron \cite{Cox},
neutron \cite{Jirak,Yoshizawa}, and electron diffraction \cite{Mori}
experiments for $0.3\le x\le 0.5$.
They show that the charge modulation for these compositions in the CO state
is 1:1 as
observed in La$_{0.5}$Ca$_{0.5}$MnO$_{3}$ \cite
{Chen,Radaelli} and Pr$_{0.5}$Ca$_{0.5}$MnO$_{3}$\cite{Mori}. These
observations illustrate the very sensitive balance between the CO and FMM
phases as $x \rightarrow 0.3$ for materials with small $<r_A>$ as in Pr$%
_{1-x}$Ca$_x$MnO$_3$. This was clearly demonstrated in the recent work on La$%
_{\frac{5}{8}-x}$Pr$_x$Ca$_{\frac{3}{8}}$MnO$_3$ ~\cite{cheong}. These authors
showed that by changing $x$ i.e. the relative amounts of La and Pr, the
ground state could be changed from an FMM state (for $x=0$) to the CO state
(for $x=%
\frac{5}{8}$). For intermediate values of $x$, $T_C$ was lowered on
increasing $x$. In fact for a range of temperatures a micron scale phase
separation (PS) between FMM and CO phases, was observed ~\cite{cheong}. The
small magnetic
fields ($\sim$ 4 T) needed to melt the CO state in Pr$_{0.7}$Ca$_{0.3}$MnO$_3
$ also brings out the delicate balance between the FMM and CO phases as $x
\rightarrow 0.3$. Therefore, the pertinent questions are: (1) What is the nature of the
metallic state obtained on melting the CO state by a magnetic field? (2) Is
it different from the low temperature FMM state in other manganites like La$_{0.67}$Ca$%
_{0.33}$MnO$_3$? We expect that low temperature specific heat experiments,
carrying the
information about principal excitations, migh provide
an answer to these questions.

In this paper we report a specific heat study of the CO state in
Pr$_{1-x}$Ca$_x$MnO$_3$ ($0.5 \ge x \ge 0.3$) and the FMM state obtained on
``melting" this CO state
with a magnetic field.
Previously, we found an anomalous excess specific heat $C^{\prime }$ of
non-magnetic
origin	in the CO state of La$_{0.5}$Ca$_{0.5}$MnO$_{3}$. Here we find that
this anomalous contribution is also present
in Pr$_{1-x}$Ca$_x$MnO$_3$ ($0.3\leq x \leq 0.5$),
even in magnetic field sufficient to induce a CO insulator to FMM
transition,
This indicates the coexistence of metallic and CO regions
(i.e. electronic phase separation). We also found that Pr$_{1-x}$Ca$_x$MnO%
$_3$ ($0.3\leq x \leq 0.5$) exhibits a large linear in temperature
contribution to the specific heat due to charge and magnetic disorder. This
linear term
is significantly reduced 
\begin{figure}[tbp]
\centerline{
\psfig{figure=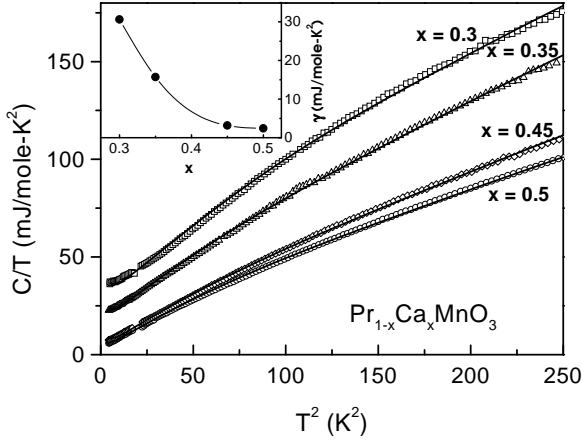,width=8.5cm,height=6.4cm,clip=}
}
\caption{Specific heat of Pr$_{1-x}$Ca$_{x}$MnO$_{3}$ samples plotted as $%
C/T $ vs $T^{2}$. Lines are fits described in text. Inset shows $\protect%
\gamma $ values for different concentrations $x$ (solid line is a guide to
the eye).}
\label{fig1}
\end{figure}
upon the application of the magnetic field due to
reduction of the disorder.




For this study we used ceramic samples of Pr$_{1-x}$Ca$_x$MnO$_3$ (x=0.3,
0.35, and 0.45), Pr$_{0.5}$Ca$_{0.5}$Mn$_{0.97}$Cr$_{0.03}$O$_3$ and La$%
_{0.5}$Ca$_{0.5}$MnO$_3$, and a single crystal of Pr$_{0.5}$Ca$_{0.5}$MnO$_3$%
. Ceramic samples were prepared by a standard solid state reaction technique
(details were described in Ref.~ \cite{Katsufuji}). X-ray powder diffraction
showed that all samples are single phase and good quality. The single
crystal was grown by the floating zone technique. The specific heat was
measured in the temperature range 2-17 K and magnetic field range 0-8.5~T by
relaxation calorimetry. This method has a relative accuracy of $\pm 3$\%.
Magnetization was measured with a commercial SQUID magnetometer, and
resistivity was measured by a standard four-probe technique.

Figure 1 shows the low temperature specific heat of the Pr$_{1-x}$Ca$_x$MnO$%
_3$ system plotted as $C/T$ vs. $T^2$, for different values of $x$ in the
temperature range from 2 to 16 K. Two anomalous contributions are evident
from these data: an excess specific heat (indicated by the nonlinearity of the
$C/T$ vs. $T^2$ plot) and the presence of a
large $\gamma T$ term in insulating $x=0.3$ and 0.35 compositions.
The low
temperature specific heat $C(T)$ of an AFM insulator should be $\beta T^{3}$,
because both lattice and AFM spin-wave contributions are proportional to $T^3$
\cite{Gopal}, which should be a straight line passing through the origin in the 
$C/T$ vs. $T^2$ plot. This is not the case for any of our samples.
Therefore, we express the low temperature
specific heat of our samples in the following form:
\begin{equation}
C = \alpha T^{-2} + \gamma T + \beta T^3 + C^{\prime}(T)
\end{equation}
The first term in Eq. (1) is
the hyperfine contribution
caused by splitting of nuclear magnetic levels of Mn and Pr ions in the field
of unpaired electrons, which was observed previously in the manganites
\cite{Woodfield,Phillips}. The origin
\begin{figure}[tbp]
\centerline{
\psfig{figure=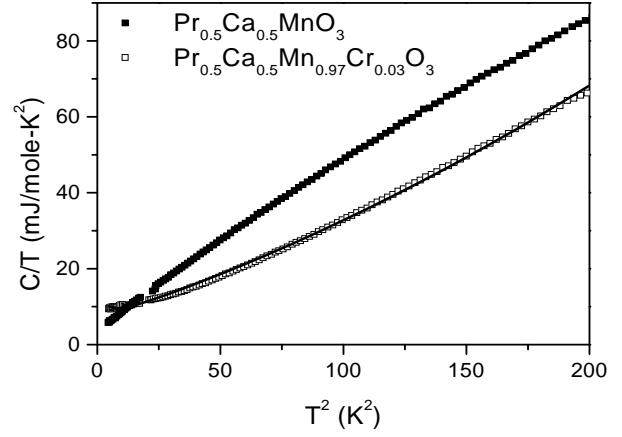,width=8.5cm,height=6.2cm,clip=}
}
\caption{Specific heat of charge-ordered Pr$_{0.5}$Ca$_{0.5}$MnO$_{3}$ and
ferromagnetic metallic Pr$_{0.5}$Ca$_{0.5}$Mn$_{0.97}$Cr$_{0.03}$O$_{3}$
samples plotted as $C/T$ vs $T^{2}$. Line is fit described in text.}
\label{fig2}
\end{figure}
of the $\gamma T$ for these electrically insulating samples will be
discussed later.
We discovered the anomalous contribution $C^{\prime}(T)$ previously in
charge-ordered La$_{1-x}$Ca$_x$MnO$_3$ ($x\approx 0.5$) ~\cite{vera}. The
temperature dependence of this contribution corresponds to non-magnetic
excitations with dispersion relation $\epsilon = \Delta + Bq^2$, where $%
\Delta $ is an energy gap and $q$ is a wave vector ~\cite{vera}.
Since Pr$_{1-x}$Ca$_x$MnO$_3$ ($0.3 \leq x \leq 0.5$) has the same charge
modulation as in La$_{0.5}$Ca$_{0.5}$MnO$_{3}$ and Pr$_{0.5}$Ca$_{0.5}$MnO$%
_{3}$ \cite{Jirak,Yoshizawa,Cox,Mori}, one possible origin of the $%
C^{\prime}(T)$ contribution is the presence of low frequency optical phonons 
\cite{vera} corresponding to an out of phase motion of Mn$^{3+}$ and Mn$%
^{4+} $ ions in their respective planes in the structure. Another possible
origin is orbital excitations ~\cite{vera}, since the CO in manganites is
accompanied by an orientational ordering of the $d_{z^2}$ orbitals of Mn$%
^{3+}$. Clearly, other experiments are needed to determine the origin of the
$C^{\prime}(T)$ term.

The results of fitting the data to Eq. (1) are shown in Table 1. Fits
corresponding to values listed in Table 1 are shown in Fig.~1 as solid
lines. The value of $\alpha$ is larger than that  reported in Ref.~
\cite{Woodfield} for the Mn hyperfine term and it decreases with $x$. This
suggests that the hyperfine term is a result of contributions
from Mn and Pr nuclei.
However, lower temperature measurements are needed to determine more
precisely the hyperfine contribution of these materials. The $\beta T^3$
term is larger for Pr$_{1-x} $Ca$_{x}$MnO$_3$ than for charge-ordered La$%
_{1-x}$Ca$_{x}$MnO$_3$ ($x\approx 0.5$) \cite{vera} which indicates a
smaller Debye temperature and/or a larger antiferromagnetic spin-wave contribution
($\propto T^3$) in Pr$_{1-x}$Ca$_{x} $MnO$_3$. The data for these all
charge-ordered samples can be best fit without the higher-order lattice term,
$\beta_5 T^5 $.
We note that values of $\Delta$ which differ
from those listed in Table 1 by $\pm 0.2$ meV also give a good fit with a
change of the other parameters by 10-15\%. This is the error limit for the
fit.

To confirm that the $C^{\prime }$ contribution is present only in the CO
state, we compare the specific heat of 
\begin{figure}[tbp]
\centerline{
\psfig{figure=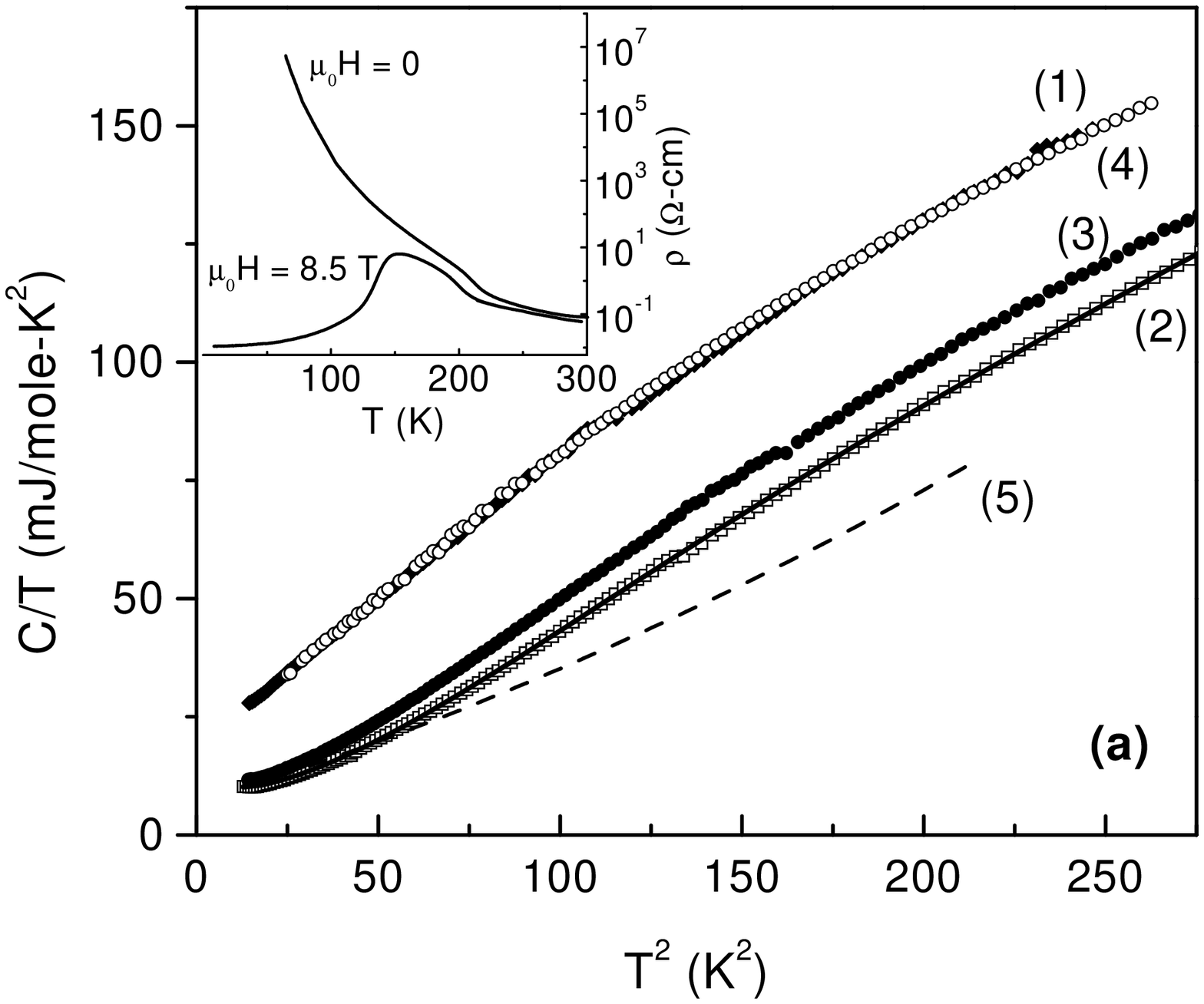,width=8.5cm,height=7.1cm,clip=}
}
\centerline{
\psfig{figure=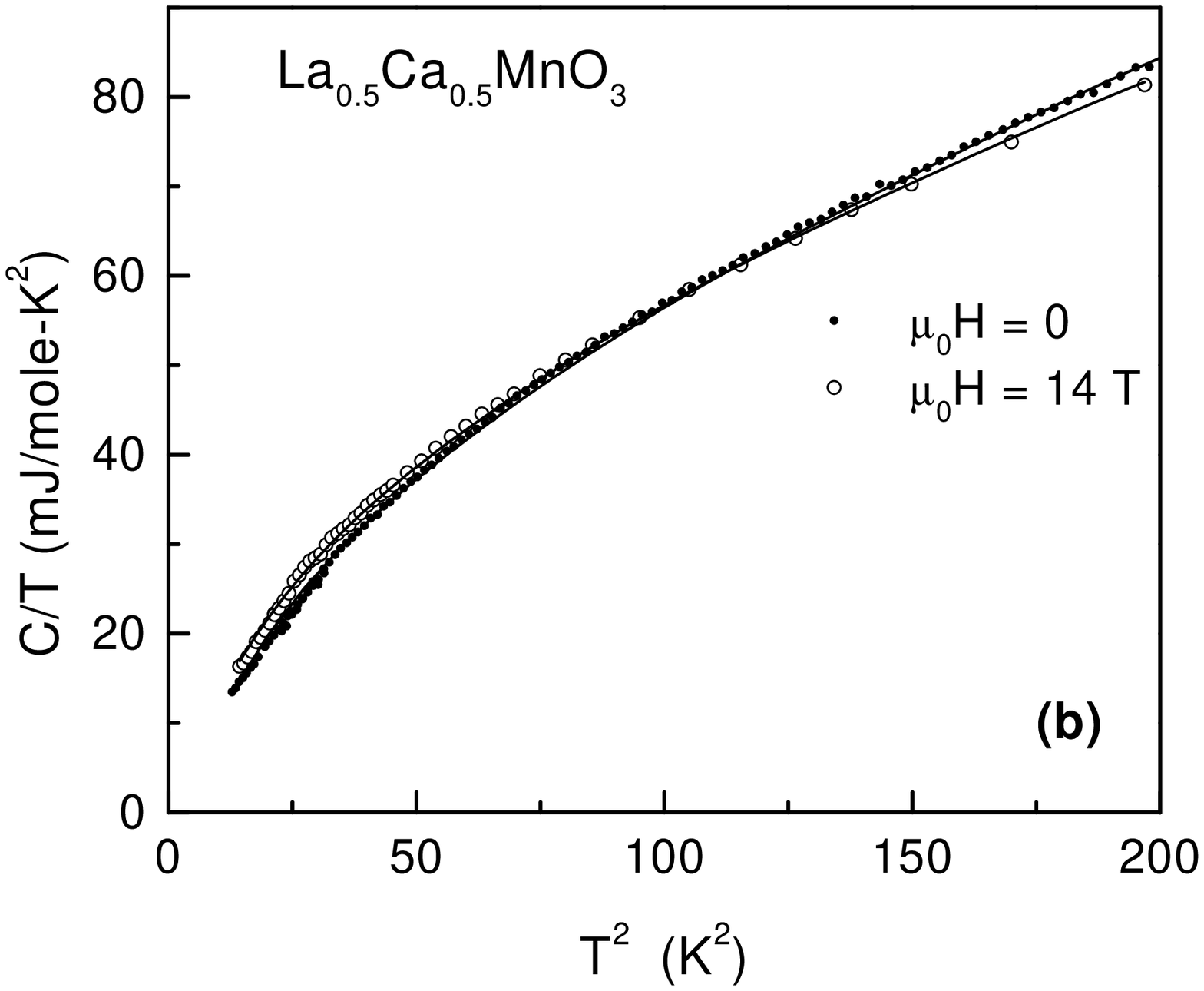,width=8.5cm,height=6.7cm,clip=}
}
\caption{(a)  Specific heat of charge-ordered Pr$_{0.65}$Ca$_{0.35}$MnO$_{3}$
sample in different magnetic fields: (1) diamonds - $\protect\mu _{0}H=0$;
(2) open squares - $\protect\mu _{0}H=8.5$ T; (3) filled circles - $\protect%
\mu _{0}H=2$~T (field was reduced to 2 T after application of 8.5 T field);
(4) open circles - $\protect\mu _{0}H=2$ T (zero field cooled sample).
Dashed line (5) representing specific heat data of FMM Pr$_{0.5}$Ca$_{0.5}$Mn%
$_{0.97}$Cr$_{0.03}$O$_{3}$ sample is shown for comparison. Solid line is
fit to $\protect\mu _{0}H=8.5$ T data (described in text). The inset shows
the temperature dependence of the resistivity of Pr$_{0.65}$Ca$_{0.35}$MnO$%
_{3}$ in zero and 8.5 T magnetic field. (b) Specific heat of charge-ordered
La$_{0.5}$Ca$_{0.5}$MnO$_{3}$ with and without magnetic field. Lines are
fits described in text.}
\label{fig3}
\end{figure}
Pr$_{0.5}$Ca$_{0.5}$MnO$_{3}$ and Pr$%
_{0.5}$Ca$_{0.5}$Mn$_{0.97}$Cr$_{0.03}$O$_{3}$ (Fig. 2). Doping of Cr at the
Mn site destroys the CO in this compound and results in a FM metallic state
at low temperature ~\cite{Katsufuji}. We fit the Pr$_{0.5}$Ca$_{0.5}$Mn$%
_{0.97}$Cr$_{0.03}$O$_{3}$ data to the form $C=\alpha T^{-2}+\gamma T+\beta
T^{3}+\beta _{5}T^{5}+\delta T^{3/2}$, where $\gamma T$ is a charge carrier
contribution and $\delta T^{3/2}$ is a ferromagnetic spin-wave contribution.
Values of the fitting parameters are listed in Table 1. The best fit also
requires $\beta _{5}T^{5}$ with $\beta _{5}=0.46\pm 1$ $\mu $J/mole-K$^{6}$
and $\delta =0$. As was noted in previous work \cite{Hamilton,Lees,Ghivelder}%
, it is difficult to resolve the ferromagnetic spin-wave contribution to the
specific heat in FMM manganites due to its small value and the presence of
the $\gamma T$ contribution. The charge carrier  contribution is
close to that found in other FMM manganites \cite{Lees,Hamilton}. The large
excess specific heat in the charge-ordered Pr$_{0.5}$Ca$_{0.5}$MnO$_{3}$
sample compared to the FMM Pr$_{0.5}$Ca$_{0.5}$Mn$_{0.97}$Cr$_{0.03}$O$_{3}$%
, is very evident in Fig. 2.

We now return to the unexpected $\gamma T$ term in the specific heat of
these electrically insulating samples. The inset in figure 1 shows a plot of 
$\gamma $ values for the different $x$. As $x$ approaches 0.3, the values of 
$\gamma $ become increasingly large. In fact the $\gamma $ values are much
larger than observed in metallic manganites \cite
{Woodfield,Lees,Hamilton,Ghivelder}, where $\gamma $ values were found to be
in the range 3 - 7 mJ/mole K$^{2}$. As $x$ changes from 0.5 to 0.3, the
magnetic ordering at low temperatures changes from CE type to pseudo CE type
magnetic ordering and FM correlations increase in the system ~\cite
{Jirak,Yoshizawa}. This leads to frustration in the spin system and to a
spin-glass behavior, which was observed in neutron scattering \cite
{Yoshizawa} and AC susceptibility measurements ~\cite{maignan}. 
This spin glass behavior is a possible cause for the large values of $\gamma $
for $x=0.3$ and $x=0.35$ samples. A large linear in temperature contribution
to the specific heat was observed in many spin glasses previously \cite{cg}.
Recently a large $\gamma T$ term
associated with spin disorder was also found in insulating LaMnO$_{3+\delta }
$ \cite{GhivelderI}.

Another possible cause for the presence of the large $\gamma T$ term is
charge disorder. The charge modulation in Pr$_{1-x}$Ca$_{x}$MnO$_{3}$ ($%
0.3\leq x<0.5$) is the same as for $x=0.5$ \cite{Yoshizawa}, but for $x<0.5$
there are not enough Mn$^{4+}$ ions to provide a perfect 1:1 charge ordering
of Mn$^{4+}$ and Mn$^{3+}$, resulting in charge disorder for $x<0.5$. The
two level states of different charge configurations would also have a linear
in temperature specific heat similar to spin glasses (or glasses \cite
{PhillipsII}). We believe that the spin and charge disorder is responsible
for the large values of $\gamma $ for $x=0.3$ and $x=0.35$ samples.


Next we discuss the effect of a magnetic field.
Figure 3a shows $C/T$ vs. $T^{2}$
for the $x=0.35$ sample at different fields and different thermal and
magnetic history. Resistivity measurements (Fig. 3a, inset) show that in a
magnetic field of 8.5 T the material is in the metallic state. Magnetization
measurements show that a magnetic field of 6 T is sufficient to induce a
transition from the AFM CO insulating state to the FM metallic state for our
$x=0.35$ sample. We find that the specific heat of
the Pr$_{0.65}$Ca$_{0.35}$MnO$_{3}$ sample decreases dramatically in a
magnetic field of 8.5 T (Fig. 3a). Moreover, the specific heat exhibits a
memory effect characteristic of the CO manganites \cite{Kuwahara}. After
reducing the magnetic field from 8.5 T to 2 T (while not increasing $T$
above 20 K), the sample tends to retain its smaller specific heat, while if
a 2 T magnetic field is applied to the zero field cooled (ZFC) sample (not
subjected previously to a magnetic field sufficient to ``melt'' the charge
ordering), the specific heat is exactly the same as for zero field (Fig.
3a). This behavior indicates that the decrease of the specific heat in 8.5 T
is associated with the ``melting'' of the charge ordering in this system.

A comparison of specific heat of Pr$_{0.65}$Ca$_{0.35}$MnO$_{3}$ in magnetic
field of 8.5 T and the FMM Pr$_{0.5}$Ca$_{0.5}$Mn$_{0.97}$Cr$_{0.03}$O$_{3}$
sample shows that the excess specific heat, $C^{\prime }$, does not
disappear completely in a 8.5 T. Rather, it appears to move to a higher
temperature (Fig.~3a). 
Since the $C^{\prime }$ contribution is found only in the CO state, our data
suggest that charge ordering is not destroyed completely by 8.5 T magnetic
field. This result is surprising, since the resistivity and magnetization
indicate that the CO is ``melted'', yielding a metallic FM state.
However, our results agree with neutron scattering studies \cite{Yoshizawa}
that
suggest coexistence of metallic and CO phases in a magnetic field. Our
results are even more striking for La$_{0.5}$Ca$_{0.5}$MnO$_3$ in a ``melting"
magnetic field of 14 T (Fig. 3b and Table 1). Although a small
$\gamma T$ term appears
in this field indicating   the presence of charge carriers, the $C^{\prime}
$ term remains essentially the same. These results indicate a coexistence of
FMM and CO phases, an electronic phase separation, in a field above the
``melting"
magnetic field. In addition, since Pr$_{0.65}$Ca$_{0.35}$MnO$_{3}$ is in the
FM state at 8.5 T, the presence of the $C^{\prime }$ contribution in our 8.5
T data confirms that $C^{\prime }$ is not of AFM origin, as proposed in
Ref.~\cite{Lees}.

The lattice, charge carrier, hyperfine,
FM spin wave and $C^{\prime }$ terms were included in the fit of the
magnetic field data. The fitting results are listed in Table 1. The
variation of the fitting parameters within 5\% of the cited values could
still lead to a reasonable fit, but larger variations of these parameters
lead to considerable deviations from the experimental data.

As we discussed above, the magnetic state of Pr$_{1-x}$Ca$_{x}$MnO$_{3}$
becomes more disordered as $x$ decreases towards 0.3 and exhibits spin glass
behavior. We expect the magnetic disorder to be reduced when a magnetic
field induces a transition to the FMM (well ordered) state. Indeed, we
observe a large decrease of $\gamma $ in a magnetic field of 8.5 T (Fig. 3,
Table 1) to a value found for metallic manganites
\cite{Woodfield,Hamilton,Lees,Ghivelder}. The change in magnetic state
appeares to have removed (at least partially) the
magnetic disorder responsible for the anomalously large $\gamma T$ term in
insulating Pr$_{0.65}$Ca$_{0.35}$MnO$_{3}$.
However, it is not possible to
determine what part of the $\gamma T$ term found at 8.5 T corresponds to the
charge carrier contribution since some magnetic disordere may remain.

The $\beta $ value, $0.12\pm 0.01$ mJ/mole-K$^{4}$, is
close to the $\beta $ value found in metallic manganites. This decrease of $%
\beta $ in magnetic field is likely due to two effects: the absence of the
AFM spin-wave contribution $\beta _{{\rm AFM}}T^{3}$ at this magnetic field
and the decrease of the lattice contribution $\beta _{{\rm latt}}T^{3}$ due
to the decrease of the unit cell volume at the ``melting'' field \cite
{striction}. The 8.5 T data fit well without including a FM spin-wave
contribution. The change of the specific heat in the FMM state (difference
between data sets (3) and (2) in Fig. 3) does not correspond to a
temperature and magnetic field dependence of FM spin-waves, indicating, that
other contributions to $C$ are also changing in the magnetic field. This
does not permit us to resolve a FM spin-wave contribution from our data.



We have shown that an anomalous $C^{\prime}$ term is observed
for the composition range studied ($0.3<x<0.5$). These results support
our view that this contribution caused by low frequency
excitations due to the arrangement of the Mn$^{+3}$ and Mn$^{4+}$ ions in
separate sublattices in the CO state. TEM studies have shown that the charge
modulation in the CO state for $0.3<x<0.5$ is 1:1, i.e. the same as in the $%
x=0.5$ compound. This is supported by the observation that the gap value in
the $C^{\prime}$ excitation spectrum is similar for all the compounds
studied here.

In conclusion, we have found a large linear in $T$ term in the low
temperature specific heat of Pr$_{1-x}$Ca$_{x}$MnO$_{3}$ as $x$ approaches
the AFM-FM boundary of the phase diagram. This contribution is most likely
associated with spin and charge disorder. In the FM state induced by a
magnetic field the $\gamma T$ contribution decreases to the typical value
for metallic manganites. We found that Pr$_{1-x}$Ca$_{x}$MnO$_{3}$ ($0.3\leq
x\leq 0.5$) compounds, which have the same type of the charge ordering as La$%
_{1-x}$Ca$_{x}$MnO$_{3}$ ($x\approx 0.5$), have an excess specific heat, $%
C^{\prime }$, of non-magnetic origin. A magnetic field sufficient to induce
the transition from the insulating AFM to the metallic FM state in the Pr$%
_{0.65}$Ca$_{0.35}$MnO$_{3}$ and La$_{0.5}$Ca$_{0.5}$MnO$_{3}$ compounds
modifies, but does not eliminate the $C^{\prime }$ contribution. This suggest
that charge ordering is not completely destroyed by the ``melting" magnetic
field and CO and metallic regions coexist in the sample.


\noindent {\bf Acknowledgment}: We thank A. J. Millis for helpful
discussion, H. D. Drew, J. Cerne, M. Grayson, J. Simpson, G. Jenkins
and D. Schmadel for use of 14 T magnet. This work is supported in part by
the NSF-MRSEC at Maryland, NSF-DMR-9802513 at Rutgers.

\newpage 
\begin{table}[tbp]
\caption{Summary of the fitting results for the specific heat data.
The units of different quantities are:
$\alpha$ (mJ-K/mole),
$\gamma$  (mJ/mole-K$^2$),  $\beta$  (mJ/mole-K$^4$),  $\Delta$  (meV),
and $B$  (meV-\AA$^2$).}
\begin{center}
\begin{tabular}{||c|c|c|c|c|c||}
x & $\alpha$  & $\gamma$  & $\beta$ 
& $\Delta$  & $B$  \\
(Pr$_{1-x}$Ca$_x$MnO$_3$) &  &  &  &  & \\ \hline
0.3 & 63 & 30.6 & 0.30 & 1.73 & 12.2 \\ \hline
0.35 & 56 & 15.7 & 0.39 & 1.15 & 20.7 \\ \hline
0.45 & 28 & 3.1 & 0.31 & 1.15 & 24.2 \\ \hline
0.5 (0\% Cr) & 22 & 2.4 & 0.26 & 1.15 & 23.7 \\ \hline
0.5 (3\% Cr) & 26 & 6.5 & 0.22 &  &  \\ \hline
0.35 ($\mu _0 H = 8.5$ T) & 66 & 7.0 & 0.12 & 3.46 & 7.6 \\ \hline
La$_{0.5}$Ca$_{0.5}$MnO$_3$ &  &  &  &	& \\
($\mu _0 H = 0$) &  &  & 0.14 & 0.72 & 17 \\ \hline
La$_{0.5}$Ca$_{0.5}$MnO$_3$ &  &  &  &	& \\
($\mu _0 H = 14$ T) &  & 2.2 & 0.11 & 0.72 & 16.9 \\ 
\end{tabular}
\end{center}
\end{table}

\end{document}